\documentclass[sigconf]{acmart}
\AtBeginDocument{%
  }

\usepackage{microtype}                 
\usepackage{textcomp}                  
\usepackage{multirow}   
\usepackage{enumitem}
\usepackage{amsmath}

\copyrightyear{2026}
\acmYear{2026}
\setcopyright{cc}
\setcctype{by}
\acmConference[VRST '26]{32nd ACM Symposium on Virtual Reality Software and Technology}{November 16--18, 2026}{Sendai, Japan}
\acmBooktitle{32nd ACM Symposium on Virtual Reality Software and Technology (VRST '26), November 16--18, 2026, Sendai, Japan}
\acmDOI{10.1145/3822517.3848661}
\acmISBN{979-8-4007-2811-2/2026/11}

\graphicspath{{figures/}{pictures/}{Images/}{./}} 

\acmSubmissionID{1084}

\begin{document}

\title{Remote Expert Evaluation of Tabletop Avatar Gaze Control in Asymmetric Multi-Scale MR--VR Collaboration}

\author{Carlos Mosquera}
\email{carlos.mosquera\_villanueva@siemens.com}
\affiliation{%
  \institution{Siemens A.G.}
  \city{Munich}
  \country{Germany}
}
\affiliation{%
  \institution{Graz University of Technology}
  \city{Graz}
  \country{Austria}
}

\author{Neven Elsayed}
\email{nelsayed@know-center.at}
\affiliation{%
  \institution{Know-Center GmbH}
  \city{Graz}
  \country{Austria}
}

\author{Ernst Kruijff}
\email{ernst.kruijff@h-brs.de}
\affiliation{%
  \institution{Bonn-Rhein-Sieg University of Applied Sciences}
  \city{Sankt Augustin}
  \country{Germany}
}

\author{Joseph Newman}
\email{joseph.newman@siemens.com}
\affiliation{%
  \institution{Siemens A.G.}
  \city{Munich}
  \country{Germany}
}

\author{Valeriya Lehrbaum}
\email{valeriya.lehrbaum@siemens.com}
\affiliation{%
  \institution{Siemens A.G.}
  \city{Munich}
  \country{Germany}
}

\author{Eduardo Veas}
\email{eveas@tugraz.at}
\affiliation{%
  \institution{Graz University of Technology}
  \city{Graz}
  \country{Austria}
}

\renewcommand{\shortauthors}{Mosquera et al.}

\begin{abstract}
This paper proposes an integrated setup for a multi-scale asymmetric collaboration scenario in the context of a "remote-expert" use case, focusing on the expert side. In this system, the remote expert (with an exocentric view in Mixed Reality (MR)), uses a Tangible User Interface (TUI) to precisely position and rotate their avatar within a scaled, tabletop World-in-Miniature (WIM). Inside the WIM, the expert’s avatar and the operator share the same scale and egocentric view, while the expert retains a global tabletop overview of the plant to support collaboration.

Interviews with industry experts during the participatory design phase revealed the need for an egocentric view inside the plant on the remote-expert side to enable full collaboration. We designed and study three interaction-visualization combinations to integrate and control the expert avatar's gaze: VR (fully immersive with direct head control), Head (floating screen with direct head control), and Touch (on-avatar screen with touchpad control).

We evaluated these methods with 29 participants across two fault-detection tasks: Same Machine, where both collaborators inspect the same machine, and Twin Machine, where they inspect identical machines in separate locations. The VR condition, which combines immersion, direct head control, and a full-scale view, yielded the strongest overall performance and user experience, while the tabletop avatar setup was well received for collaboration. Our findings inform the design of gaze-control techniques for asymmetric MR--VR collaboration and reveal challenges and opportunities for future work.
\end{abstract}



\begin{teaserfigure}
 \includegraphics[width=\textwidth]{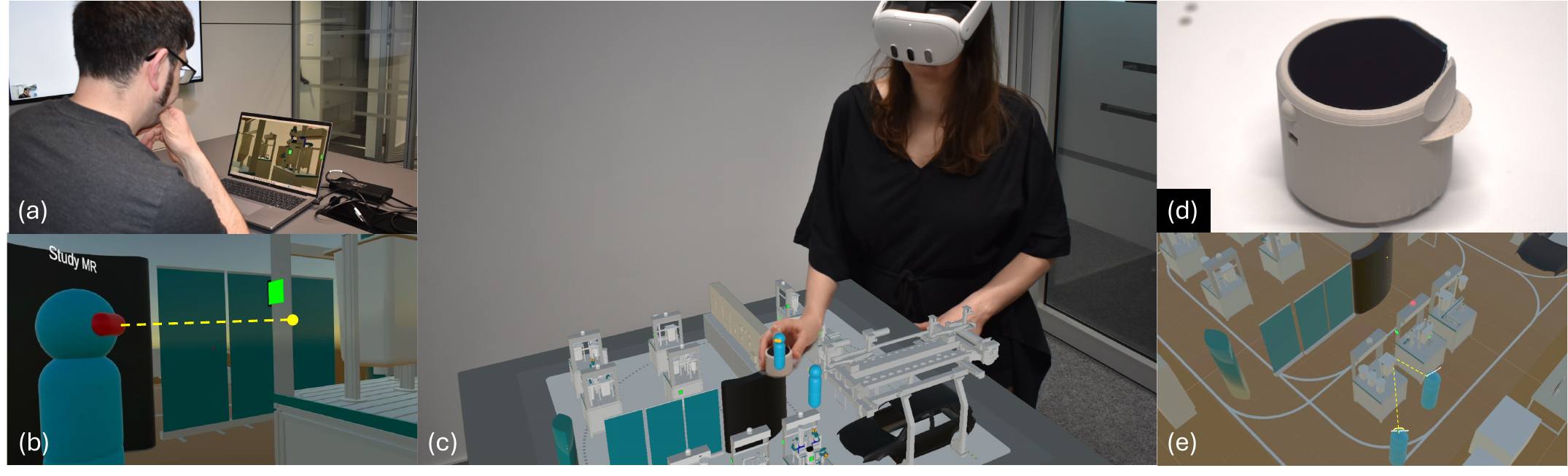}
  \caption{Asymmetric MR--VR Collaboration (remote-expert use case): (a) Operator using Desktop Virtual Reality (VR). (b) Operator's egocentric VR view. (c) Remote expert using Mixed Reality (MR) interacting with their avatar in the World-in-Miniature (WIM) plant. (d) Tangible prop with integrated touchpad for avatar control. (e) Remote expert's exocentric view.}
  \Description{Five-panel composite photo showing the asymmetric collaboration setup. Panel (a) shows an operator seated at a desktop with a computer using desktop VR. Panel (b) shows the operator's first-person VR view inside a virtual industrial plant. Panel (c) shows a remote expert wearing an MR headset, reaching toward a miniature tabletop plant model with small avatars visible inside it. Panel (d) shows a close-up of a handheld tangible prop with an integrated touchpad. Panel (e) shows an exocentric third-person view of the remote expert standing at the tabletop.}
  \label{fig:teaser}
\end{teaserfigure}


\maketitle

\section{Introduction}
Collaboration in Extended Reality (XR) is moving towards scenarios where individuals with heterogeneous hardware and disparate roles work together across the reality-virtuality continuum, giving rise to \emph{asymmetric} setups~\cite{assaf_cues_2024,EnsCollab19}. This is made increasingly convenient by consumer XR video see-through Head Mounted Displays (HMDs)~\cite{MetaQuest3,AppleVisionPro,ViveFocusVision}, which let each collaborator move between fully immersive Virtual Reality (VR) and Mixed Reality (MR) or Augmented Reality (AR) and adopt the mode best suited to their role. This is particularly relevant in \textit{Remote Expert} scenarios, the most common form of asymmetric collaboration, which typically pair a local operator with a remote knowledgeable person~\cite{sereno_collaborative_2022}. Such support can also span differences in \emph{scale}, as in multiscale setups like the Giant-Ant~\cite{chenechal_when_2016,MixedSpaceCollab05}, where a \textit{Giant}  (Director) with an AR exocentric view guides an \textit{Ant}(Actor) with a VR egocentric view through complex environments such as a navigational maze~\cite{MixedSpaceCollab05} or a co-manipulation task~\cite{chenechal_when_2016}.
A common enabling metaphor is the World-in-Miniature (WIM), a 3D replica that functions as both map and input space \cite{danyluk_design_2021}. A tabletop WIM lets the expert retain an exocentric plant overview while the immersed operator works at 1:1 scale \cite{assaf_cues_2024,chenechal_when_2016}. These asymmetric viewpoints, however, create the  \textit{viewpoint gap} \cite{PiumsomboonAwareness19}: workspace awareness degrades as perspectives diverge \cite{assaf_cues_2024}, and conveying the expert point of attention within the operator’s immersive view remains a key design challenge \cite{sereno_collaborative_2022}.

Through interviews with industry experts (see Section~\ref{sec:industrial-use-case}), we found that remote experts need an in-plant egocentric view in parallel with the exocentric tabletop overview to collaborate effectively. Yet, there is no established way to integrate and control this expert-centric egocentric view. In this work, we systematically explore this open design space, focusing on the expert side.
We present an asymmetric MR--VR setup for remote industrial fault detection, integrating existing techniques---tabletop WIM, TUI control, and workspace awareness (WA) cues---into a coherent system for empirical evaluation. We adopt TUIs because they ground avatar manipulation on the physical tabletop, providing the stability and spatial precision that mid-air gestures lack~\cite{mosquera_study_2025}, while co-locating control with the miniature plant the expert is inspecting. Our system empowers a Giant expert, using an MR-HMD, to manipulate their mini avatar within a tabletop WIM via a tangible prop. By physically controlling the avatar’s position and rotation, the expert establishes spatial presence in the operator’s (VR user) workspace. To enhance situational awareness, the expert can observe their avatar’s egocentric view through a floating virtual screen or by transitioning into a full egocentric perspective. How best to visualize and control this view remains an open question.

To answer that, we evaluated three interaction–visualization combinations for controlling the avatar’s gaze in an industrial fault detection context. In a user study with 29 participants acting as remote experts, we measured performance and experience across two collaborative tasks: (1) \emph{Same Machine}, where both collaborators work on the same machine, and (2) \emph{Twin Machine}, where they inspect identical machines in different locations. Our goal was to determine how MR point‑of‑view and gaze‑control techniques affect performance, user experience, and collaboration in asymmetric MR–VR remote‑expert work, and to derive design implications.

Our contributions are: (1) An asymmetric MR--VR remote-expert setup using a tabletop WIM and tangible avatar control in a multi-scale configuration. (2) Three interaction-visualization techniques (VR, Head, Touch) for controlling and presenting the expert avatar’s gaze. (3) A user study with 29 participants across two fault-detection tasks, yielding design implications for asymmetric MR--VR remote-expert collaboration.

\section{Related Work}
This research builds on prior work in asymmetric MR collaboration, WA, tabletop TUIs, and WIM. We review these areas to ground our mixed-space scenario, where a tabletop MR expert uses a tangible prop to control their avatar position, rotation and gaze.

\subsection{Asymmetric Collaboration in Mixed Reality}
In recent years, research in Collaborative Virtual Environments (CVEs) has shifted from symmetric interaction toward asymmetric viewpoints, scales, and interfaces to support distinct user roles. A major milestone in this direction was the transitional interface enabling seamless switching between Shared AR and Shared VR Environments for co-located collaboration \cite{Kiyokawa99}. In parallel, the MagicBook allowed users to move along the reality--virtuality continuum, inspecting an exocentric AR scene on a physical book or ``flying inside'' for an egocentric VR view \cite{MagicBook01}. Together, these systems demonstrated how mixed AR/VR collaborators can maintain mutual awareness through visual metaphors — AR users seeing VR partners as miniature avatars, and VR users perceiving AR partners as large virtual heads above the scene.

One of the earliest rigorous evaluations of mixed-space collaboration showed that pairing an AR exocentric \emph{Director} with a VR egocentric \emph{Actor} improved performance over individual work and was comparable to VR-VR collaboration \cite{MixedSpaceCollab05}. The study also highlighted that misalignment between control and visual spaces harms performance by increasing the mental effort of coordinate transformations \cite{MixedSpaceCollab05}. This exocentric--egocentric paradigm was later extended into the Giant--Ant configuration for concurrent object manipulation, where the Giant provides coarse global translations and the Ant performs fine in-situ adjustments \cite{chenechal_when_2016}.

Another exocentric-egocentric scenario is the \textit{god-like interaction} metaphor, which links indoor tabletop interfaces with outdoor AR users \cite{Stafford06}. In this setup, an expert maintains a global overview and guides the immersed user via a \textit{giant hand} that appears to descend from the sky \cite{Stafford06}. This concept was later adapted for architectural design in Dollhouse VR, which introduced a "see-through ceiling" that projects the tabletop user’s face into the virtual space to support social awareness and eye contact \cite{DollhouseVR15,DollhouseVREval18}.

Recent systems leverage AR-VR role asymmetries and tangible interaction to support guidance across different scales \cite{PiumsomboonCOVAR17,PiumsomboonGiant19,PiumsomboonMiniMe18}. In GMC, an AR \emph{Giant} can physically steer a VR \emph{Miniature}'s viewpoint via a TUI \cite{PiumsomboonGiant19}. Snow Dome adds recursive scaling while preserving a shared reference frame \cite{PiumsomboonSnowDome18}.
These asymmetric setups make it difficult to understand a partner’s focus and perspective across radically different viewpoints (viewpoint gap)\cite{assaf_cues_2024, EnsCollab19}. To bridge this gap, researchers have focused on workspace awareness cues.

\subsection{Workspace Awareness (WA)}
WA is a real-time understanding of how others are interacting with a shared workspace---who is present, where they are working, and what they are doing \cite{gutwin_workspace_1996,gutwin_descriptive_2002}. While it often emerges naturally in co-located work through visible actions and their effects, digital and distributed settings reduce perceptual access and weaken nonverbal cues, which can undermine WA \cite{gutwin_workspace_1996,gutwin_descriptive_2002}. In 3D collaborative virtual environments, systems therefore add explicit awareness cues (e.g., gaze rays and view cones) to convey attention and field of view (FOV) and to mitigate the XR viewpoint gap \cite{gutwin_descriptive_2002,dyck_awareness_nodate,assaf_cues_2024}.

In MR collaboration, HMDs reduce eye contact and facial-expression cues, weakening social and WA \cite{augello_design_2023,gutwin_workspace_1996,gutwin_descriptive_2002}. To compensate, designers add explicit awareness cues and metaphors---from global orientation aids to local, over-the-shoulder views---to communicate where a partner is looking and what they can see \cite{provenzano_designing_2007}. In asymmetric MR--VR setups, 3D pointing metaphors (e.g., 2D pointer/3D ray) and viewpoint cues such as FoV frustums and head-gaze rays help bridge the XR viewpoint gap and improve coordination \cite{duval_improving_2014,assaf_cues_2024,PiumsomboonAwareness19}.

\subsection{Tabletop Tangible User Interfaces}
TUIs couple digital information to physical objects and environments, leveraging physical affordances to make interaction more understandable \cite{ishii_tangible_1997,fishkin_taxonomy_2004}. A central benefit is graspable object control, where users manipulate virtual content through physical \textit{bricks}  that provide direct, embodied control \cite{fitzmaurice_bricks_1995,ishii_tangible_1997}. In tabletop settings, graspable tangibles support spatial tasks for maps/landmarks \cite{ishii_tangible_1997} and for physically grounded object manipulation \cite{kato_magiccup_2003}. Recent work further indicates that tabletop-grounded tangibles (e.g., knob-shaped proxies) can outperform mid-air gestures by exploiting the table for stability and ergonomic support \cite{mosquera_study_2025}.

In collaborative settings, tabletop interfaces support shared discussion and workspace awareness \cite{park_context_nodate,kolkmeier_little_2018}. In asymmetric collaboration, actuated tokens and independent viewpoints can improve co-presence and task performance by making remote intent and activity visible on the shared table \cite{park_context_nodate,lee_physical-virtual_2018}. HoloBots \cite{ihara_holobots_2023} extends this approach by coupling holographic telepresence with gaze-driven tabletop robots as physical surrogates, enabling miniature-body interaction and supporting intent communication.

\subsection{World in Miniature (WIM)}
Although our system uses a real tabletop MR setup, the WIM metaphor remains central \cite{stoakley_virtual_1995,pausch_navigation_1995}. WIMs present an environment as a manipulable miniature that supports overview, navigation, and at\mbox{-}a\mbox{-}distance manipulation \cite{coffey_interactive_2012,danyluk_design_2021}. Early work showed that tangible props (e.g., clipboards) can anchor perception and ease transitions between miniature and life\mbox{-}size views \cite{stoakley_virtual_1995,danyluk_design_2021}. Later systems improve WIM travel with pre\mbox{-}oriented teleportation and viewpoint previews, reducing mental rotation and disorientation \cite{elvezio_travel_2017,weissker_try_2024}.

Prior work shows the value of complementing exocentric and egocentric viewpoints in asymmetric mixed-space collaboration, yet the viewpoint gap still limits WA. Recent work highlights visual WA cues (e.g., FoV frustums and head-gaze rays) as key for coordinating attention in MR. However, TUIs for direct avatar control in tabletop WIMs remain underexplored, and their combination with WA cues to jointly control avatar position and gaze remains unevaluated. We address this gap by evaluating such a scenario.

\begin{figure*}
    \centering
    \includegraphics[width=1\linewidth]{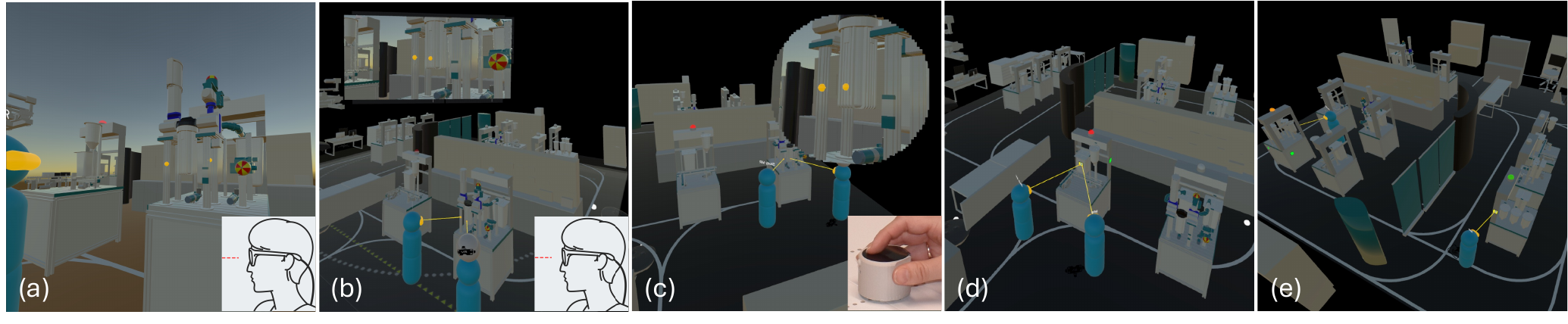}
    \caption{Interaction Methods: (a) VR: Fully immersive, direct head control, (b) Head: Floating screen, direct head control, (c) Touch: Small on-top avatar screen, touch control. Tasks: (d) Task 1 (T1): Same Machine, (e) Task 2 (T2): Twin Machine.}
    \Description{Five-panel illustration showing three interaction methods and two tasks. Panels (a)--(c) depict the VR, Head, and Touch conditions respectively, each showing the expert's viewpoint and control mechanism. Panel (d) shows Task 1 where the expert and operator share the same machine. Panel (e) shows Task 2 where they work on twin machines side by side.}
    \label{fig:AllMethods}
\end{figure*}

\section{System Design}
We use the industrial remote-expert scenario as our guiding use case; findings may generalize to other collaborative contexts.

\subsection{Industrial Use Case}\label{sec:industrial-use-case}
We focus on the remote expert use case, which is among the most popular scenarios for XR collaboration~\cite{EnsCollab19}. In this scenario, a knowledgeable person guides a local operator through a physical task, such as industrial fault detection. This setup supports asymmetric roles: the expert provides guidance and oversight, while the operator performs hands-on work in the real environment.

To gain deeper insights into the remote expert fault detection context, through a participatory design process we conducted unstructured interviews with three XR industry experts. From these discussions, we derived several requirements for the expert role. First, the expert needs to maintain a global overview of the entire plant to effectively coordinate with an operator focused on a specific area or machine. Second, for close inspection, the expert needs to see from the same perspective as the operator in order to understand their viewpoint and local context. Third, real-world scenarios may involve multiple experts and operators working together, requiring support for multi-party collaboration.

Based on the analysis of the fault detection workflow, we modeled the task as comprising two main phases: an initial navigation and wayfinding phase, in which the expert guides the operator from their current location to the target machine, followed by a joint inspection and repair phase, in which both collaborators closely inspect the machine, identify the fault, and coordinate the repair action.
Other domains may benefit similarly from this asymmetric MR–VR collaboration setup, such as plant design, where on‑site operators can be directly involved in layout and process discussions, or emergency response scenarios, where a remote expert can guide and teleoperate a robot deployed in the affected area.

\subsection{Tabletop Avatar Design} \label{sec:tabletop-avatar}
A key component is a tabletop with an augmented, scaled WIM plant in MR. This gives the expert a persistent global overview, addressing the first requirement. We chose a tabletop specifically because it provides a stable physical surface for TUI-based avatar control~\cite{mosquera_study_2025}, spatially anchors the WIM so that positions on the miniature map directly to plant locations, and lets multiple co-located experts gather around a shared reference for face-to-face coordination~\cite{cheng_tabletop_2015,MuellerARVRCollab19}. Prior work has used tabletop WIMs for asymmetric collaboration~\cite{Stafford06,DollhouseVR15}.

Another requirement is that the expert can see from the operator's perspective to correctly interpret the local situation and provide grounded instructions. Prior work on collaborative XR has extensively analyzed such egocentric, view-aligned configurations—such as shared first-person views and overlaid annotations—where both operator and expert inhabit the same scale and perspective in a virtual or augmented scene \cite{oda_virtual_2015,wang_you_2018,kolkmeier_little_2018}. In our case, however, this egocentric requirement is in tension with an exocentric tabletop overview of the plant, which is valuable for planning, navigation, and maintaining workspace awareness \cite{sereno_collaborative_2022,assaf_cues_2024}.

Several systems have explored transitions between exocentric and egocentric perspectives in XR, allowing users to move between an overview miniature and an immersive first-person view \cite{Kiyokawa99,MagicBook01,MagicBookRev23}. While these transitions can support spatial understanding, they expose a drawback for our scenario: when the expert is fully immersed in an egocentric view, they temporarily lose direct access to the global tabletop representation needed for coordination and navigation across distant parts of the plant. At the same time, the task’s navigation phase requires the expert to direct the operator to new locations, where prior work has relied on exocentric \textit{god hand} metaphors or tangible props to indicate paths and targets from above the scene \cite{underkoffler_urp_1999,ihara_holobots_2023,lee_physical-virtual_2018}.

We replace the disembodied god hand with an embodied expert avatar in the tabletop WIM. The avatar is a controllable proxy: the expert positions it anywhere in the WIM to indicate locations and paths while retaining the exocentric overview. When needed, the expert triggers a transition to the avatar's viewpoint, entering the scene through their miniature self. This has two advantages. First, the expert controls where they transition. Second, in egocentric view both users are co-located at equal scale, supporting mutual understanding and first-person guidance~\cite{genay_being_2022,nguyen_user_2020}. The avatar thus mediates between overview and immersion.

\subsection{Tangible Prop Design}
We control the expert avatar with a tangible prop, as TUIs provide rich physical affordances and precise control, especially on a tabletop surface, where they improve ergonomics and precision for virtual object manipulation ~\cite{mosquera_study_2025}. Following design guidelines for compact, hand-scale tabletop tangibles, we gave the prop a rounded, compact form factor with a cylindrical shape that enhances controllability and fine-grained adjustments through its graspable form and continuous rotational affordances~\cite{voelker_cylindrical}, letting the expert simultaneously control their avatar's position and orientation.

For tracking, we embedded a Sony Toio~\cite{SonyToio} swarm robot, providing compact 2D tabletop tracking~\cite{nakagaki_hermits_2020}. To extend interaction beyond pure manipulation, we integrated a rounded touchscreen driven by an ESP32-based board~\cite{esp32lcd} and a portable battery, enabling self-contained visual feedback and touch input. Side handles improve grip while keeping the touchscreen unobstructed; on-device buttons offer additional input channels. The resulting prop weighs 56g, with a diameter of 45mm and height of 39mm, as shown in Fig.~\ref{fig:teaser}(d).

\subsection{Workspace Awareness Components}

Following recent WA surveys~\cite{assaf_cues_2024}, we designed the system to support the main WA questions (who, where, and where-looking) through visual cues. WA informs the system design; the study evaluates how the expert visualizes and controls these cues, not WA itself.

For \emph{Who} (presence/identity), each user is represented by a full‑body avatar with a distinctive color and a name label rendered above the head. This makes it easy to identify collaborators at a glance and attribute actions to specific users.

For \emph{Where} (location), both expert and operator can directly control their avatars position in the shared MR workspace, keeping movement and spatial arrangements visible to all collaborators.

For \emph{Where} (gaze/view), rather than a full frustum, we used a head-gaze indicator, in line with findings that subtle gaze cues are less intrusive yet effective for WA \cite{provenzano_designing_2007,wang_you_2018}. We cast a ray from the avatar's head along its viewing direction, rendering a thin line that ends in a small hit sphere where it meets the environment, continuously indicating each collaborator's focus of attention (Figure ~\ref{fig:teaser}b).

\subsection{Apparatus And Collaboration Server}
The expert used the Quest 3 MR headset, a standalone device with full-color passthrough, operated via two 6-DoF controllers or hand tracking~\cite{MetaQuest3}. The operator used a desktop PC.
For networking, we used a cloud-hosted WebSocket server to synchronize avatars and head poses in real time, inspired by existing frameworks ~\cite{netcodegameobject}.\\

The system combines a tabletop MR setup, a tangible prop for avatar control, and explicit WA cues including a head-gaze ray, making it crucial that the expert can perceive their avatar's view and control its gaze. Finding no prior work on combined view-and-control techniques for a miniature avatar, we conducted a pilot to inform the technique design.

\section{Interaction Design Pilot Phase}
Effectively using the head-gaze cue depends on two aspects: \emph{how} the avatar's view is visualized and \emph{how} the expert controls it. Three industry experts (Section~\ref{sec:industrial-use-case}) participated in this pilot to evaluate alternatives and finalize the methods for evaluation.

\subsection{Viewpoint Representation Alternatives}
Prior work has provided egocentric views of tabletop scenes by transitioning from an exocentric tabletop view to full VR~\cite{MagicBook01,MagicBookRev23}. 

While intuitive, letting users experience the scene from within-it has drawbacks in our case: switching to VR removes the exocentric plant overview, can reduce spatial orientation~\cite{FeldWeyersCityTour21}, and raises cognitive load from the mode switch and context change~\cite{MixedSpaceCollab05}.
To address this, we draw inspiration from WIM research, where preview windows show the target viewpoint before teleportation to reduce disorientation~\cite{elvezio_travel_2017,huang_preview_2025}. We adapt this to our tabletop setting: instead of transitioning the expert fully into VR, a dedicated display shows what the miniature avatar sees, letting the expert inspect and adjust its viewpoint while preserving the exocentric overview.
The size, shape, and placement of such preview windows form a coupled design space rather than independent factors. In our pilot, we therefore explored two holistic alternatives. First, a small screen on the tangible prop, compact to limit occlusion~\cite{elvezio_travel_2017}. Second, a larger preview around the tabletop that dynamically follows the HMD~\cite{huang_preview_2025,coffey_interactive_2012}. We iteratively refined both with domain experts.

Based on their feedback, we converged on three final visualizations. The fully VR view needed little adjustment, presenting a 1:1 egocentric mapping of the operator's perspective. The medium preview is a floating rectangular screen (~$1.0 \times 0.5$m), positioned on the opposite side of the tabletop to stay visible as the expert moves. The compact preview is a small circular screen (~$30$cm diameter) mounted on the tangible prop that tilts to face the expert, improving visibility while minimizing occlusion.

\subsection{Control Alternatives}
This dimension concerns how the expert controls the avatar's gaze. Given the visualization alternatives above, full VR lends itself naturally to \emph{direct head mapping}, where the expert's head movements directly define the avatar's viewing direction.

For preview windows, beyond \emph{direct head} mapping, we explored \emph{indirect} head control, where the avatar's gaze depends on the head's pose relative to the screen---like a portal---so moving around it shifts the viewpoint, letting the expert ``peek'' into the WIM without fully entering VR~\cite{elvezio_travel_2017,huang_preview_2025,coffey_interactive_2012}. In addition, we investigated \emph{external} control via input on the tangible prop itself, such as a touchpad on top, keeping control co-located with the avatar proxy and consistent with our tangible design. We excluded mid-air gestures and detached devices, as they cannot be cleanly integrated into the prop and conflict with our goal of a compact, self-contained TUI.

In expert evaluation, indirect head control, though initially promising, caused confusion as its mapping felt unnatural, while the prop-mounted touchpad was valued for keeping the device compact. We thus narrowed the design to two control methods: direct head and touchpad control. These two control methods were then paired with the three visualizations to form the final conditions (Section~\ref{sec:final-interaction-design}).

\subsection{Final Interaction Design}\label{sec:final-interaction-design}
Based on expert feedback, we paired each visualization with a gaze control method. The fully VR visualization used its natural direct head control (\emph{VR} condition). The floating rectangular preview used direct head control, preferred over touch here (\emph{Head} condition). The small circular on-top screen used touchpad control, as experts found a preview spatially and functionally coupled with the touch surface easier to focus on (\emph{Touch} condition).

All conditions shared the same initial sequence: the expert positioned and oriented the avatar on the WIM plant, then pressed and held the prop against the tabletop via the handles to enable gaze control. This maintains continuous contact with the prop: this is vital in VR, where the user cannot see the tabletop or prop and would otherwise struggle to relocate it.
Once gaze control is enabled, each condition proceeds differently:

\subsubsection{VR Condition}
After pressing the button, the expert transitions into fully immersive VR at the avatar's current position, controlling its gaze naturally through head movements for a first-person view into the plant. A hitsphere at the head-gaze ray's endpoint marks the expert's focus, enabling precise control and clear communication to the operator.
 Figure \ref{fig:AllMethods}a illustrates this condition.

\subsubsection{Head Condition}
After pressing the button, the expert remains in the MR tabletop environment, seeing the avatar's egocentric view on the large rectangular preview screen, with its gaze controlled directly by head movements. A hitsphere at the head-gaze ray's endpoint indicates where he is looking, communicating his focus to the operator while still retaining the exocentric tabletop overview. Figure \ref{fig:AllMethods}b illustrates this condition.

\subsubsection{Touch Condition}
As in \emph{Head}, the expert stays in the MR tabletop environment, but the avatar's egocentric view now appears on the small circular screen mounted on the tangible, with its head-gaze controlled via the integrated touchpad. This lets the expert keep the tabletop overview while simultaneously monitoring the on-prop egocentric view. Unlike Head, where attention often shifts toward the distant large screen, this better supports continuous awareness of both the miniature plant and the avatar's egocentric view. Figure \ref{fig:AllMethods}c illustrates this condition.\\

The three conditions trade off egocentric view size against parallel tabletop visibility. \emph{VR} offers a full-scale, immersive first-person view but temporarily loses direct tabletop access. \emph{Head} reduces it to a large but bounded preview while restoring continuous access to the miniature plant and all avatars' head-gaze rays. \emph{Touch} shrinks it to a small on-prop screen but maximizes parallel monitoring of the tabletop and avatar viewpoint, giving strong awareness of both global context and local detail.

\section{Study Design}

\subsection{Collaboration Task Design}
In designing the collaboration tasks, we grounded our choices in the industrial remote expert requirements outlined in Section~\ref{sec:industrial-use-case} and in prior work on asymmetric MR collaboration and navigation support~\cite{MixedSpaceCollab05,Stafford06,PiumsomboonGiant19}. The aim was to create a compact, repeatable scenario that captures the main phases of a typical remote fault detection session while stressing the need for effective gaze communication between expert and operator.

The first phase addresses \emph{navigation and wayfinding}. Using the exocentric view of the tabletop plant WIM, the remote expert identifies the target machine by spotting status indicators that may be difficult to notice from the operator perspective. The expert then guides the immersed operator to the machine. Rather than relying only on spoken turn-by-turn directions as in audio-only support~\cite{MixedSpaceCollab05}, the expert uses their "mini" avatar as a landmark, as in the "god-hand" \cite{Stafford06}. By moving the avatar along the route and placing it near the destination, the expert highlights the path and key reference points, which the operator sees at full (1:1) scale.

The second phase centers on \emph{joint machine close inspection}. Once the operator reaches the desired machine, both collaborators inspect it iteratively to identify the faulty component. This tightly coupled work increases the need for precise reference to specific parts and clear gaze communication. The task therefore directly exercises the expert's gaze-control methods, requiring the expert to position the avatar at the machine and steer its gaze with one of the three interaction techniques to coordinate attention with the operator.

\subsubsection{Inspection Puzzle}
To evaluate non-expert participants in a fault-detection-inspired context, we designed a simplified machine inspection puzzle rather than a real industrial one. Our goal was to preserve the core collaboration characteristics of remote fault detection---joint inspection, mutual reference to parts, and agreement on the faulty component---while keeping the task easy to learn. We drew inspiration from prior collaborative research tasks~\cite{PiumsomboonAwareness19}, where two users inspect objects and must align their gazes on the same target for a passcode to be revealed, then verbally confirm whether their values match.

Adapting this idea, we implemented a sticker-solver puzzle. Once both users were co-located, several unrevealed stickers appeared on the machine's sides, each ( $10\,\times\,10$\,cm) a potential fault indicator. Placing the gaze hit-sphere on a sticker revealed a passcode: the expert saw a number, the operator a letter. The pair verbally checked these against a printed index-to-letter sheet held by the expert (e.g., 1$\rightarrow$a, 2$\rightarrow$b, etc). If no match, they tried other stickers until finding the correct one. The operator then selected it to trigger the repair and ``fix'' the machine. This mechanic promotes active gaze use and verbal negotiation while keeping the task simple.

\subsubsection{Fault Detection Tasks}
We defined two tasks sharing the same goal---finding the machine fault---but differing in the collaborators' spatial arrangement. Both use the inspection puzzle described above as their close-inspection mechanic. The first is a well-studied configuration where both users jointly analyze the same machine; the second places them at different plant locations. This second design emerged from expert input, where certain diagnostics compare a functional reference machine with a faulty one to localize the problem, demanding greater parallel attention to the exocentric overview than the first.

\textit{Task~1 (T1): Same Machine.} Both expert and operator collaborate on the same machine. From the tabletop overview, the expert identifies the faulty machine by its red status indicator and guides the operator there. Once the operator enables it, the inspection puzzle activates: stickers appear around the machine, and the collaborators coordinate their head-gaze on the same stickers to reveal the candidate number-letter combination. After jointly identifying the correct sticker, the operator selects it to complete the repair. Figure \ref{fig:AllMethods}d illustrates this task.

\textit{Task~2 (T2): Twin Machine.} The expert and operator work on two spatially separated but identical machines: a faulty one and a reference. From the tabletop overview, the expert identifies both via status lights, guides the operator to the faulty (orange) machine, and moves their avatar to the reference (green) one. After the operator enables the faulty machine, the inspection puzzle activates on both: matching marks appear. To reveal a candidate number-letter combination, each collaborator looks at the same corresponding mark on their own machine; a single gaze reveals the code (the expert sees a number, the operator a letter). They then verbally determine which mark yields a matching pair per the predefined mapping, and once identified, the operator selects it to repair the faulty machine. Figure \ref{fig:AllMethods}e illustrates this task.

\subsection{Study Procedure}
Sessions took place in a quiet room. After welcoming the participant, the supervisor explained the study's purpose, obtained informed consent, and gave a short overview of the setup, tasks, inspection puzzle, and gaze interaction methods. Participants then completed a pre-questionnaire on demographics and prior MR/VR experience. The supervisor adjusted the table height, configured the MR environment by positioning a virtual screen and aligning the virtual plant via QR-code tracking, and fitted the HMD for comfort.


A short two-part training followed: participants first practiced all interaction methods in a dedicated scene, then rehearsed both fault detection tasks with a virtual collaborator, including navigation and the sticker-based inspection puzzle. In the actual study, participants took the role of the remote expert, communicating via a Teams call with the operator—a study supervisor who followed a standardized script, acting as a naive operator led by the expert and providing only neutral acknowledgments to ensure consistent behavior across conditions—and completed 36 trials, with short breaks available on request. Afterward, the system automatically stored the data, and the participant completed a post-study questionnaire on their experience and subjective impressions. Including briefing, training, and questionnaires, each session lasted approximately 1 hour and 20 minutes.

All procedures for this study were approved by the TU Graz Ethics Committee (GZ: EK-102/2025).

\subsection{Experiment}
For the experiment, we evaluated the three avatar head-gaze interaction methods introduced in Section~\ref{sec:final-interaction-design} (VR, Head, Touch) across the two fault detection tasks described above (T1: Same Machine and T2: Twin Machine).
To control puzzle behavior, we defined two difficulty levels based on the number of incorrect stickers that had to be inspected before the solution became available: at level~1, the first inspected sticker was always incorrect and the next one was the solution; at level~2, the first two inspected stickers were incorrect and the solution appeared only on a later sticker. The system tracked inspected stickers and dynamically set the solution position accordingly.

We used a within-subjects design in which all participants, acting as remote experts, experienced all interaction methods, both tasks, and both difficulty levels. Conditions were counterbalanced using a Latin square distribution to reduce order effects. In total, each participant completed $2$~tasks $\times$ $3$~interaction methods $\times$ $2$~difficulty levels $\times$ $3$~repetitions, resulting in $36$~trials per participant. Because each condition combines a specific visualization, display, and control method, the study compares bundled interface configurations rather than isolated factors; causal attribution to individual dimensions is not possible.

\subsection{Hypotheses}
We formulated our hypotheses to compare the 3 avatar's gaze control methods in terms of task performance, perceived workload, and user preference, as well as the overall effectiveness of the tabletop avatar control for remote collaboration.

\textbf{H1 (T1 – Same Machine).} For T1, which focuses on detailed inspection around a single machine, we expect the VR condition to outperform Head and Touch. Alignment between interaction and visualization spaces has been shown to reduce mental effort and improve performance~\cite{MixedSpaceCollab05}, and our pilot (Section~\ref{sec:final-interaction-design}) indicated that direct head control in an immersive view felt most natural for precise targeting. We therefore hypothesize that VR will yield faster task completion, lower perceived workload, and higher user preference.

\textbf{H2 (T2 – Twin Machine).} For T2, which requires parallel attention to two spatially separated machines, we expect Head and Touch to outperform VR. As discussed in Section~\ref{sec:tabletop-avatar}, full immersion temporarily removes direct access to the exocentric tabletop overview needed for cross-machine coordination. Because Head and Touch preserve continuous tabletop visibility while T2 demands monitoring both machines, we hypothesize they will yield faster completion, lower workload, and higher preference than VR.

\textbf{H3 (Overall tabletop avatar control).} Across all conditions, we expect the tabletop avatar control to support effective remote collaboration. Tabletop tangibles provide stability and precision for spatial control~\cite{mosquera_study_2025}, and physical tokens linked to virtual agents have been shown to increase co-presence~\cite{lee_physical-virtual_2018}. Because our avatar combines tangible manipulation with workspace awareness cues on a shared tabletop, we hypothesize that participants will rate it as effective for collaboration and coordination, and report high levels of presence and preference for the remote-expert role.

\subsection{Measures}

For the quantitative data, the system logged per-trial timestamps to compute three metrics:


\begin{itemize}[leftmargin=*]
    \item \emph{Time Completion ($\text{solution\_time} - \text{enabling}$ $\text{\_machine\_time}$):} Since all conditions share the same navigation and wayfinding phase, timing begins when the operator enables the machine, isolating the joint inspection phase where the gaze-control methods differ.
    \item  \emph{Overshoot:} Count of tangible (re)selections, reflecting how often the expert needed to interrupt gaze control to reposition the avatar before re-engaging.
    \item \emph{Gaze Overshoot:} Count of times the gaze hit-sphere entered a sticker area, reflecting a combination of deliberate inspections and unintentional re-entries during gaze steering.
\end{itemize}

For qualitative data, participants completed a post-study questionnaire in which they rated each interaction method separately: a fast motion sickness scale, the NASA-TLX (workload), the System Usability Scale (SUS, usability and preference), and four items adapted from the IPQ (physical presence, involvement, attention to virtual plant, attention to tabletop). Further items addressed embodiment, interaction naturalness, and collaboration aspects (coordination and communication quality). Open-ended questions invited comments on each condition and on the overall the setup.

\section{Results} \label{sec:results}

\begin{figure}
    \centering
\includegraphics[width=1.0\linewidth]{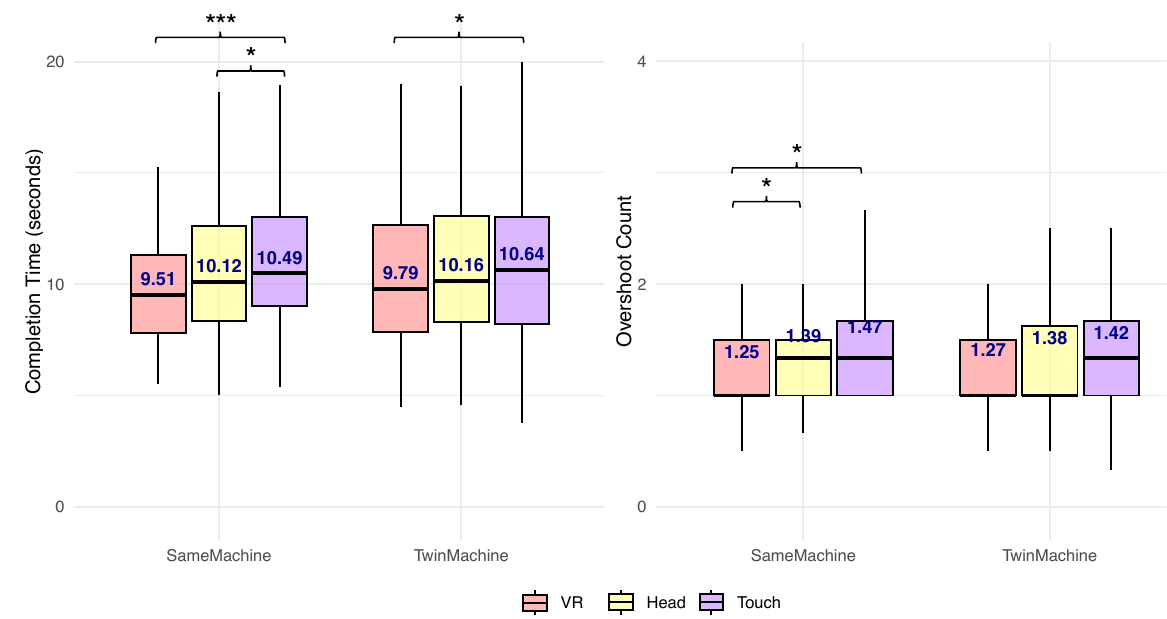}
    \caption{Pairwise results across interaction methods per task. Left: Completion Time. Right: Overshoot.}
    \Description{Two box plots with pairwise significance brackets. The left chart shows completion time per interaction method for each task. The right chart shows overshoot counts per interaction method for each task.}
    \label{fig:All_results}
\end{figure}

29 participants (18 men and 11 women) were voluntarily recruited from the company, age $M=28.9$ ($SD=5.2$). 17 participants did not wear corrective glasses, and 12 did. They rated their AR/VR experience on a 1–10 scale with $M=4.4$ ($SD=2.5$), and their HMD experience with $M=2.4$ ($SD=2.3$). Participants reported no physical or mental condition. In total, 998 of 1,044 expected trials (29 × 36) were analysed; the remainder were excluded due to one participant's recording malfunction (subjective data retained) and a few time-constrained sessions that ended early.

\subsection{Data Analysis Approach}

Shapiro–Wilk tests indicated non-normal distributions for all measures; we therefore used Friedman tests with Holm-corrected post-hoc Wilcoxon signed-rank comparisons throughout. To account for difficulty levels, completion time, overshoot, and gaze overshoot were normalized by the minimum required steps (difficulty + 1). Outliers were removed before aggregation using Z-score filtering (|z| > 3). Per-participant scores were then computed as the median across repetitions and difficulty levels for each condition, yielding one value per participant per method. An ART ANOVA revealed no significant interaction between interaction method and task type for completion time \textit{($F(2,135) = 0.64,\ p = 0.530$)}, confirming that method effects were consistent across both tasks; T1 and T2 were therefore analyzed separately. All analyses were conducted in R. Detailed statistical results, including pairwise comparisons, effect sizes, outlier exclusion counts, questionnaire items with response scales, and per-condition descriptive statistics for all subjective measures, are provided in the supplementary material.

\begin{figure}
    \centering
    \includegraphics[width=1.1\linewidth]{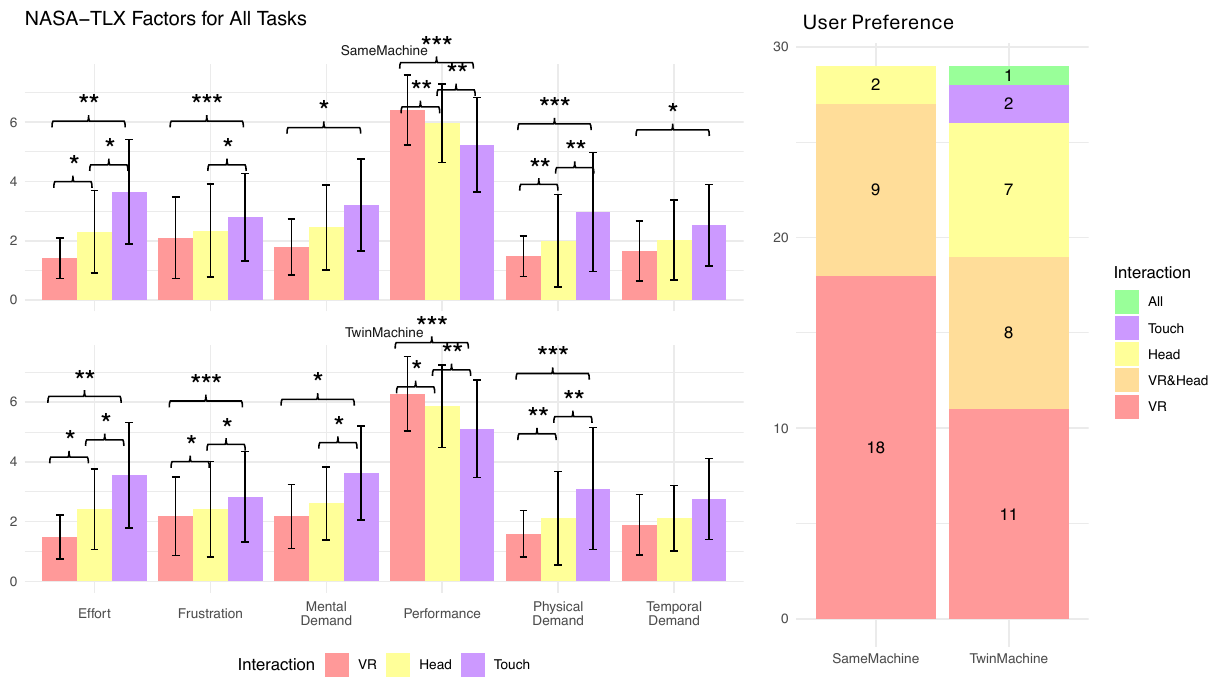}
    \caption{Subjective results. Left: NASA--TLX subscales by method and task. Right: Method preference by task.}
    \Description{Three charts arranged side by side. The left two charts show box plots of six NASA--TLX subscales for each interaction method, with Task 1 on top and Task 2 on the bottom. The right chart shows stacked bar plots of method preference rankings for Task 1 and Task 2.}
    \label{fig:NASATLX}
\end{figure}

\subsection{T1 Same Machine}

\subsubsection{Quantitative Measures}

A Friedman test revealed significant differences in completion time across interaction methods  \textit{($\chi^2_2= 16.93, p< 0.001, W= 0.302$)}. Post-hoc Wilcoxon tests revealed that Touch took significantly longer than VR and Head, while no significant differences were found between VR and Head (Figure \ref{fig:All_results}).
A Friedman test revealed significant differences in overshoot across interaction methods  \textit{($\chi^2_2= 9.16, p= 0.01, W= 0.164$)}. Post-hoc Wilcoxon tests revealed that VR took significantly less overshoot than Touch and Head, while no significant differences were found between Touch and Head (Figure \ref{fig:All_results}).
No significant differences were found for gaze overshoot.

\subsubsection{Task load, User Choice}

For task load, Friedman tests revealed significant effects of interaction method on all NASA–TLX subscales: Mental Demand \textit{($\chi^2_2 = 14.25,\ p < .01,\ \text{W} = 0.25$)}, Physical Demand \textit{($\chi^2_2 = 34.07,\ p < .001,\ \text{W} = 0.59$)}, Temporal Demand \textit{($\chi^2_2 = 7.05,\ p < .05,\ \text{W} = 0.12$)}, Performance \textit{($\chi^2_2 = 30.13,\ p < .001,\ \text{W} = 0.52$)}, Effort \textit{($\chi^2_2 = 21.27,\ p < .001,\ \text{W} = 0.37$)}, and Frustration \textit{($\chi^2_2 = 17.15,\ p < .001,\ \text{W} = 0.30$)}.
In all subscales, VR performed better than Touch. Figure~\ref{fig:NASATLX} illustrates all results in detail.

When asked their preferred method, 18 of 29 participants chose \textit{VR}, 2  \textit{Head}, and 9 indicated both \textit{VR} and \textit{Head} as equally preferred.

\subsection{T2 Twin Machine}

\subsubsection{Quantitative Measures}

A Friedman test revealed significant differences in completion time across interaction methods  \textit{($\chi^2_2= 13.5, p= 0.0012, W= 0.241$)}. Post-hoc Wilcoxon tests revealed that Touch took significantly longer than VR; no significant differences were found between VR and Head or between Head and Touch (Figure \ref{fig:All_results}).
No significant differences were found for overshoot or gaze overshoot.

\subsubsection{Task load, User Choice}

For task load in \textit{TwinMachine}, Friedman tests again revealed significant effects of interaction method on all NASA--TLX subscales: Mental Demand \textit{($\chi^2_2 = 13.25,\ p < .01,\ \text{W} = 0.24$)}, Physical Demand \textit{($\chi^2_2 = 29.96,\ p < .001,\ \text{W} = 0.52$)}, Temporal Demand \textit{($\chi^2_2 = 6.25,\ p < .05,\ \text{W} = 0.11$)}, Performance \textit{($\chi^2_2 = 26.94,\ p < .001,\ \text{W} = 0.46$)}, Effort \textit{($\chi^2_2 = 19.16,\ p < .001,\ \text{W} = 0.34$)}, and Frustration \textit{($\chi^2_2 = 18.97,\ p < .001,\ \text{W} = 0.33$)}.
In all subscales except in Temporal Demand, VR performed better than Touch; further results for the other tasks are shown in Figure~\ref{fig:NASATLX}.

When asked to select their preferred method, 11 of 29 participants chose \textit{VR}, 7 chose \textit{Head}, 2 chose \textit{Touch}, 8 indicated both \textit{VR} and \textit{Head} as equally preferred, and 1 selected all methods.

\subsection{General interaction}

\begin{figure}
    \centering
    \includegraphics[width=0.8\linewidth]{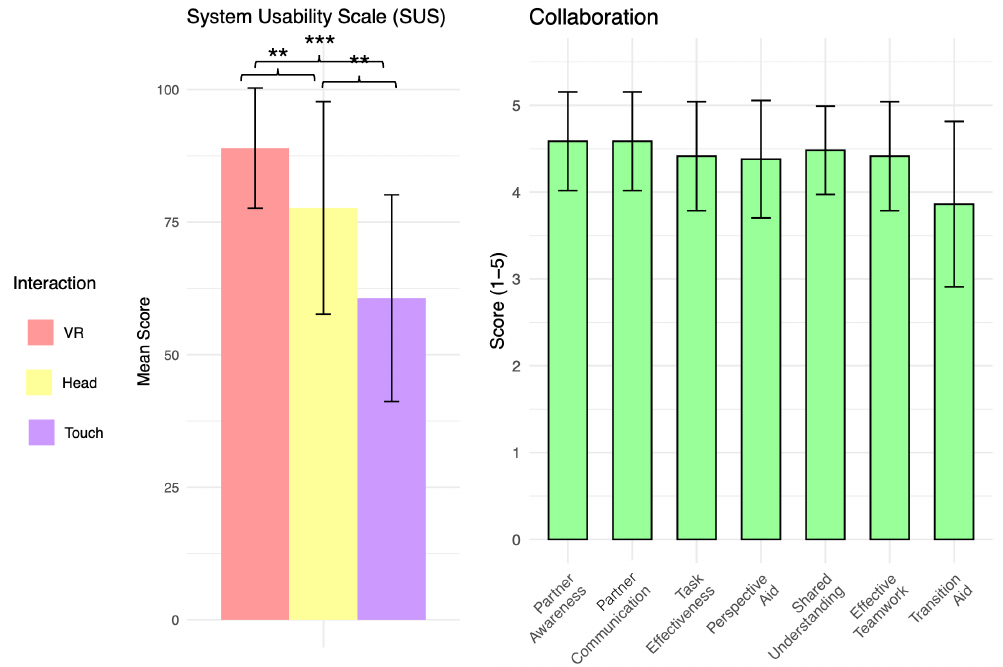}
    \caption{General subjective results. Left: System Usability Scale. Right: Overall collaboration feedback.}
    \Description{Two side-by-side charts. The left chart shows box plots of System Usability Scale scores for each interaction method with pairwise significance brackets. The right chart shows box plots of collaboration feedback questions across methods.}
    \label{fig:SUS}
\end{figure}

\subsubsection{System Usability Scale (SUS)}
For SUS scores, a Friedman test revealed a significant effect of interaction method on usability, $\chi^2_2 = 24.50,\ p < .001,\ W = 0.42$. VR showed the highest SUS scores, followed by Head and then Touch; post-hoc pairwise comparisons are shown in Figure ~\ref{fig:SUS}.

\subsubsection{Presence}
For the presence questionnaire, Friedman tests showed significant effects of interaction method on \textit{sense of physical presence} \textit{($\chi^2_2 = 33.59,\ p < .001,\ \text{W} = 0.58$)}, \textit{involvement} \textit{($\chi^2_2 = 34.03,\ p < .001,\ \text{W} = 0.59$)}, and \textit{attention to the virtual plant} \textit{($\chi^2_2 = 17.63,\ p < .001,\ \text{W} = 0.30$)}, but not on \textit{attention to the tabletop} \textit{($\chi^2_2 = 2.00,\ p = .37,\ \text{W} = 0.03$)}.
In all subscales except tabletop attention, VR and Head outperformed Touch.

\subsubsection{Embodiment}
For embodiment items, Friedman tests showed significant effects of interaction method on \textit{perceived naturalness of controlling the avatar} ($\chi^2_2 = 24.15,\ p < .001$), \textit{accuracy of avatar gaze} ($\chi^2_2 = 15.59,\ p < .001$), \textit{feeling that the avatar was an extension of the self} ($\chi^2_2 = 18.89,\ p < .001$), and \textit{naturalness and effortlessness of interacting with the virtual plant} ($\chi^2_2 = 33.32,\ p < .001$). Across all items, VR showed the highest ratings, followed by Head, and then Touch, indicating stronger embodiment in VR, intermediate embodiment with Head, and lower embodiment with Touch.

\subsubsection{General collaboration and coordination}
As for the whole system, participants reported a generally positive collaborative experience. Ratings were high for \textit{partner awareness and communication} (both $M = 4.59,\ SD = 0.57$), \textit{shared understanding} ($M = 4.48,\ SD = 0.51$), \textit{task effectiveness and effective teamwork} (both $M = 4.41,\ SD = 0.63$), and \textit{perceived benefit of the different perspectives for collaboration} ($M = 4.38,\ SD = 0.68$). The \textit{transition into first-person (avatar) view} was rated somewhat lower, though still positive ($M = 3.86,\ SD = 0.95$).

\subsection{Qualitative Feedback}

\paragraph{Head:} Participants found Head control natural, intuitive, and low effort, valuing the simultaneous view of the virtual plant and their real surroundings. Criticism focused on the avatar view sometimes falling out of sight, requiring awkward head movement or repositioning, along with the limited preview size and difficulty aligning their perspective with the avatar's.

\paragraph{Touch:} Participants found Touch easy to learn. They valued seeing both the plant and the avatar PoV on a stable screen, avoiding head/neck movement, and found it familiar, like a tablet. Criticism focused on the small view window and control area, which made it hard to see, orient, and place the gaze target, and on the non-ergonomic, fine-motor-demanding button hold during touchpad use. Some also found it slower and less intuitive, with small touches causing large, hard-to-predict view changes.

\paragraph{VR:} Participants described VR as the most natural and effortless method, with fluid control, fast targeting, strong immersion, and a vivid sense of being ``inside'' the environment. They noted it made stickers easy to see and the gaze easy to steer, improving awareness of the collaborator's focus and shared perspective. Criticism focused on full immersion hindering access to tabletop context, forcing users to exit VR to check the reference sheet or regain an overview, alongside occasional positioning issues needing large head movements or feeling ``stuck'' in awkward poses. Some also noted discomfort or motion sickness and found repeated tabletop--first-person transitions abrupt and cognitively demanding.

\paragraph{Overall Experience} Participants found the tabletop and tangible setup natural, intuitive, and smooth, with low mental and physical effort once understood. They valued the physically grounded control as comfortable and precise, and saw the system as effectively combining a global overview with fine-grained, avatar-level control to navigate efficiently while supporting partner awareness and shared understanding. Most described the experience as enjoyable and playful, noting it ``felt like playing a game'' and that they ``had a lot of fun,'' with some imagining themselves using the system in a real supervisor role.

\section{Discussion} \label{sec:discussion}

\subsection{Touch Underperformed}
Overall, Touch showed lower performance, which can be explained by participants’ feedback. Although the touchpad on the tangible prop initially seemed attractive for its compactness and co-location, it caused notable usability problems: the small control area and limited screen hindered orientation in the avatar’s egocentric view, leading to imprecise gaze targeting and higher cognitive load. Most importantly, having to press a button while simultaneously operating the touchpad introduces a dual-task demand that can negatively affect fine motor performance. Dual-task interference is known to reduce movement efficiency and increase completion time in upper-limb motor tasks due to competition for limited attentional resources \cite{Lin2015PointingDevices}. Additionally, concurrent cognitive and physical demands can increase indicators of wrist and forearm muscle fatigue during precision tasks\cite{Wagenblast2024DualTaskFatigue}. Differences in pointing-device design can further influence hand posture and forearm muscle activity, affecting comfort and perceived control.

Moreover, feedback on Touch was mixed: some participants valued its precise, "calculated" fine control, while others felt it reduced control because small movements sometimes caused large or unpredictable view changes. This split likely reflects individual differences and prior experience with input modalities, which have been shown to influence both task performance and subjective usability ratings of interaction devices \cite{ZudilovaSeinstra2010IndividualDifferences}.

\subsection{T1: Same Machine, VR strongest configuration}

In T1, VR and Head outperformed Touch in completion time, while VR showed significantly fewer re-engagements (overshoot) than both Head and Touch. VR also scored best across all NASA--TLX subscales and was preferred by the majority of participants. These results indicate that for detailed inspection around a single machine, the VR configuration provided the strongest overall performance and user experience. Because the conditions bundle visualization, display, and control method, isolating the contribution of individual factors is not possible. However, the subjective data offers converging evidence for several candidate mechanisms.
First, direct sensorimotor head-gaze mapping: both VR and Head, which share direct head control, received high general ratings for naturalness of avatar control and gaze accuracy, while Touch scored lowest on both; this aligns with prior work showing that misalignment between interaction and visualization spaces increases mental effort~\cite{MixedSpaceCollab05}. Second, a full-scale view with larger targets: in T1, VR yielded the lowest physical demand and highest self-rated performance across NASA--TLX subscales, suggesting that interacting with targets at full scale reduced motor and perceptual effort. Third, reduced split attention: general presence ratings (collected across tasks) showed that VR received significantly higher  physical presence and involvement scores, both with large effect sizes, while reported attention to the tabletop did not differ across conditions, suggesting that VR's immersion strengthened engagement without participants perceiving a loss of tabletop awareness. Fourth, stronger spatial presence: general embodiment ratings showed VR scored highest on feeling that the avatar was an extension of the self, consistent with the sense of "being there" reported in the qualitative feedback. These factors likely act in combination rather than independently. Since VR did not significantly outperform Head in completion time, H1 is partially supported: confirmed for overshoot, workload, and preference, but not for task completion time.

In addition, some participants reported brief discomfort and moments of disorientation when transitioning from the MR tabletop view into the immersive VR viewpoint and back. Prior work similarly notes that switching viewpoints and scales can increase disorientation and cognitive effort due to the abrupt context change \cite{FeldWeyersCityTour21,MuellerARVRCollab19}. In our study, this effect appeared limited: post-study Fast Motion Sickness (FMS) ratings did not indicate substantial motion sickness. Nevertheless, repeated or prolonged VR exposure could still amplify discomfort in longer sessions. To mitigate this, future designs could smooth the transition using gradual visual blending (e.g., progressive transparency or fade filters) rather than an instantaneous switch \cite{chenechal_when_2016}.

\subsection{T2: Twin Machine, VR and Head top performance}
In T2, contrary to our hypothesis (H2), Head did not outperform VR. Quantitative measures showed no significant differences, although subjective measures favored VR. This suggests that Head's parallel tabletop awareness may have provided some benefit, but did not translate into measurably better performance. As with VR, Head bundles multiple factors (a floating preview, direct head control, and continuous tabletop access), so its underperformance relative to expectations cannot be attributed to a single cause. However, participant feedback points to a specific design limitation. The system was designed for the full task workflow, including navigation, which required the user to move freely around the table. This meant that when gaze control was activated, the floating screen's initial orientation depended on where the expert happened to be standing. When participants faced the table center, the screen placement worked well; otherwise, starting head control produced awkward initial orientations, forcing uncomfortable head postures or avatar repositioning. The candidate mechanisms discussed for T1 (direct sensorimotor mapping, full-scale view, reduced split attention, and spatial presence) may have outweighed Head's parallel awareness advantage, compounded by this screen-placement issue. However, the bundled design prevents attributing Head's underperformance to any single factor.

This screen-positioning problem represents a design flaw that future work should address, either by isolating the screen-position factor to evaluate its impact independently, or by improving the dynamic positioning algorithm to adapt to the user's current orientation. Resolving this limitation would enable a fairer assessment of whether parallel tabletop awareness genuinely benefits split-attention tasks like T2.

Comparing the two tasks descriptively, the performance gap between Head and VR narrowed in T2, though the ART ANOVA found no significant method-by-task interaction. Subjective preferences showed a clearer shift: more participants chose Head or rated Head and VR equally in T2 than in T1, suggesting that Head's parallel tabletop awareness was more appealing when the task demanded attention to separated machines. Despite this subjective shift, H2 is not supported by the quantitative results.

\subsection{Positive collaboration with tabletop-tui-avatar}
Despite design flaws in Head and Touch, the tabletop avatar setup was well received: participants rated partner awareness, communication, shared understanding, and task effectiveness highly. These ratings reflect subjective experience rather than a direct measure of workspace awareness. The setup supports remote MR collaboration regardless of which gaze method is used.

The TUI prop for avatar placement and rotation was well accepted. This positive reception concerns the prop for positioning and orienting the avatar on the tabletop---a function distinct from the Touch gaze-control method, which received mixed feedback (see above). Participants reported natural, precise control, supporting our choice of a graspable, table-supported controller~\cite{ishii_tangible_1997,mosquera_study_2025}.

Additionally, our positive findings provide a basis to discuss how we can contextualize and compare our system, even without a single, clear baseline to assess performance. If we focus on the wayfinding phase, we can argue that using the tangible avatar as a persistent landmark may provide stronger spatial grounding than pure voice commands \cite{MixedSpaceCollab05}, because it augments speech with an explicit, shared spatial reference in the miniature workspace. Likewise, compared to exocentric “god-hand” pointing metaphors \cite{Stafford06}, embodying the expert as an avatar with a head-gaze ray adds an additional layer of reference: the avatar’s placement communicates where to go, while the head-gaze ray communicates what to attend to once there. Together, these cues can support a smoother transition from global navigation to local inspection by making both destinations and attentional targets visible in the shared scene.

Consequently, H3 is supported by the results.

\section{Limitations and Future Work}
The complexity of an asymmetric collaboration scenario forces design and research decisions. We focused on the expert MR tabletop user. Future work should also evaluate the operator’s perspective, including how the social presence is impacted by a collaborator with 2 representation entities (god-view and tabletop avatar).

Although our system is designed so multiple tabletop or experts can be added, we only tested a single expert and operator. Future work should explore how the system scales to multiple experts and operators, including how to manage multiple avatars and perspectives in the shared space.

Conditions were counterbalanced to reduce order effects; however, we did not analyze learning or fatigue across the 36 trials, as this was not the focus of the study. Cumulative fatigue or improved familiarity with the system may have influenced later trials. Future work could examine trial-by-trial performance trends to disentangle these effects.

Head's screen-positioning and Touch's touchpad usability emerged as design flaws that our small expert-only pilot did not catch. Future iterations should address these issues before reassessing parallel awareness. Overall, all three techniques support simultaneous exocentric and egocentric awareness, but over different spans: Touch favors exocentric context with limited egocentric detail, Head is more balanced, and VR favors egocentric detail at the cost of direct tabletop access. Users can often bridge the two views mentally and prefer whichever offers better control. Future work should validate improved designs with a larger, more diverse sample and focus either on tasks that truly demand concurrent awareness or on interfaces enabling fast, low-friction switching between views.

Finally, this setup opens new opportunities for interaction and collaboration in a space with no prior baseline. We propose our work as such a baseline, on top of which future research can explore novel interaction techniques or challenge our design decisions.

\section{Conclusion}
We presented and evaluated an integrated tabletop avatar setup for remote MR collaboration, enabling expert gaze control via three interaction-visualization conditions (VR, Head, Touch) alongside a tabletop overview. A 29-participant study showed that the VR condition, which combines immersion, direct head control, and a full-scale view, yielded the strongest performance and user experience; the tabletop avatar setup was well received for collaboration. Our findings suggest that conditions offering direct sensorimotor mapping and reduced attentional switching best support precise gaze control in asymmetric MR--VR collaboration, and reveal design challenges for this emerging space.

\bibliographystyle{ACM-Reference-Format}
\bibliography{Asymmetric-bib}

\end{document}